\documentclass[twocolumn,secnumarabic,amssymb, aps,  prd]{revtex4-1}
 \usepackage{units}
\usepackage{graphicx}	
\usepackage{amsmath}	
\usepackage{amssymb}	
 \usepackage{fancyhdr}
\usepackage[paper=a4paper
,inner=1.8cm
,outer=1.8cm
,top=3.1cm
,bottom=3.5cm]{geometry}

\newcommand{\fk}[1]{{\textcolor{black}{#1}}}		

\usepackage[
breaklinks=false,
linktocpage=false,
colorlinks=true,
linkcolor=blue,
citecolor=blue,
filecolor=blue,
pagecolor=blue,
urlcolor=blue,
frenchlinks=false,
breaklinks=true
]{hyperref}

\begin{document}

\title{In-situ analysis of optically thick nanoparticle clouds}%

\author{F. Kirchschlager${}^{1}$}
\email[]{kirchschlager@astrophysik.uni-kiel.de}
\author{S. Wolf${}^{1}$}
\author{F. Greiner${}^{2}$}
\author{S. Groth${}^{2}$}
\author{A. Labdon${}^{3}$}

\affiliation{${}^{1}$Institute of Theoretical Physics and Astrophysics, Kiel University, \fk{24105 Kiel,} Germany}
\affiliation{${}^{2}$Institute of Experimental and Applied Physics, Kiel University, \fk{24105 Kiel,} Germany}
\affiliation{${}^{3}$Department of Physics, Aberystwyth University, \fk{SY23 3FL Aberystwyth, United Kingdom}}
\date{April 13, 2017}
\begin{abstract}
\hspace*{-0.32cm}Nanoparticles grown in reactive plasmas and nanodusty plasmas gain high interest from basic science and technology. One of the great challenges of  nanodusty plasmas is the in-situ diagnostic of the nanoparticle size and refractive index. The analysis of scattered light by means of the Mie solution of the Maxwell equations was proposed and used as an in-situ size diagnostic during the past two decades. Today, imaging ellipsometry techniques and the investigation of dense, i.$\,$e. optically thick nanoparticle clouds demand for analysis methods to take multiple scattering into account.\newline
We present the first 3D Monte-Carlo polarized radiative transfer simulations of the scattered light in a dense nanodusty plasma. This technique extends the existing diagnostic methods for the in-situ analysis of the properties of nanoparticles to systems where multiple scattering can not be neglected.
\end{abstract}
\maketitle

Nanosized particles are omnipresent, be it in the atmosphere, in the  exhaust of motors or in the oceans.  Focusing on plasmas, nanoparticles are an unwanted  byproduct of fusion devices \cite{Winter1998, Virot2015} and in plasma reactors used in the semiconductor industry \cite{Selwyn1991}. Plasma aided processes are used to create  nanoparticles with unique features \cite{Kortshagen2016}, like amorphous silicon particles \cite{Wang2010}, or to embed particles in surfaces  \cite{Girshick2011}. Simulations of growth processes in reactive plasmas are on an elaborated state and give detailed insight into the growth dynamics and the spatial size distribution of the nanoparticles \cite{Agarwal2014}. All these examples have in common that they need an in-situ diagnostic for the size of the particles. Because  the refractive index of nanoparticles may not be the same as that of a large solid state system of the same material, the desired diagnostics have to estimate  both size and complex refractive index of the particles.

In a reactive plasma, using argon working gas at low pressure (some ten Pa)  with an add-on of gases like methane, acetylene, or silane,  particles can be grown from nanometer sized precursors up to micrometer sized particles. To estimate the particle size during the growth processes, Mie ellipsometry \cite{Hollenstein1994, Hayashi1994}, in particular kinetic Mie ellipsometry \cite{Groth2015}, can be applied. Advanced techniques like imaging Mie ellipsometry \cite{Greiner2012} provides full spatiotemporal information about the particle size. \fk{Such} methods are currently restricted to the assumption of single scattering. In low temperature plasmas exist two scenarios exist where this approximation fails: (a) if the particle cloud is optically thick for the light going through it and (b) in the experimental analysis of dusty plasmas where the light source for illumination is not simply a laser beam but a laser fan or a volume source.  Light sources different from laser beams are needed for imaging ellipsometry \cite{Greiner2012} or computed tomography of the particle cloud \cite{Killer2016}. Regarding atmospheric pressure microplasmas, which are used to create nanoparticles \cite{Nozaki2007}, it is supposed that the high particle number density immediately leads to high optical depths. It is clear that in such situations, each photon inside the particle cloud has to be followed until it leaves the plasma, contributing to the scattering signal when it reaches the detector. Only 3D polarized radiative transfer simulations considering multiple scattering are able to face this challenge.

In this paper we present the application of 3D multiple scattering polarized radiative transfer simulations for the analysis of dust growth in optically thick plasma clouds for the very first time.


\label{Sec_Exp}
This study is motivated by particle growth experiments \cite{Groth2015} in an argon-acetylene discharge during the accretion phase \cite{Berndt2009}. When acetylene is added to the argon plasma, amorphous hydrogenated-carbon particles (a-C:H) begin to grow. 
Due to the negative charging of the particles inside the plasma, nearly monodisperse particle growth takes place up to a few hundred nanometers. At this critical radius the particles leave the plasma confinement because the ion drag force and/or the gravitational force are larger than the confining electric field force. This also causes the formation of a void, a dust-free region in the center of the plasma \cite{Samsonov1999, Tadsen2014}.

\begin{figure} 
  \includegraphics[clip=true,width=1.0\linewidth]{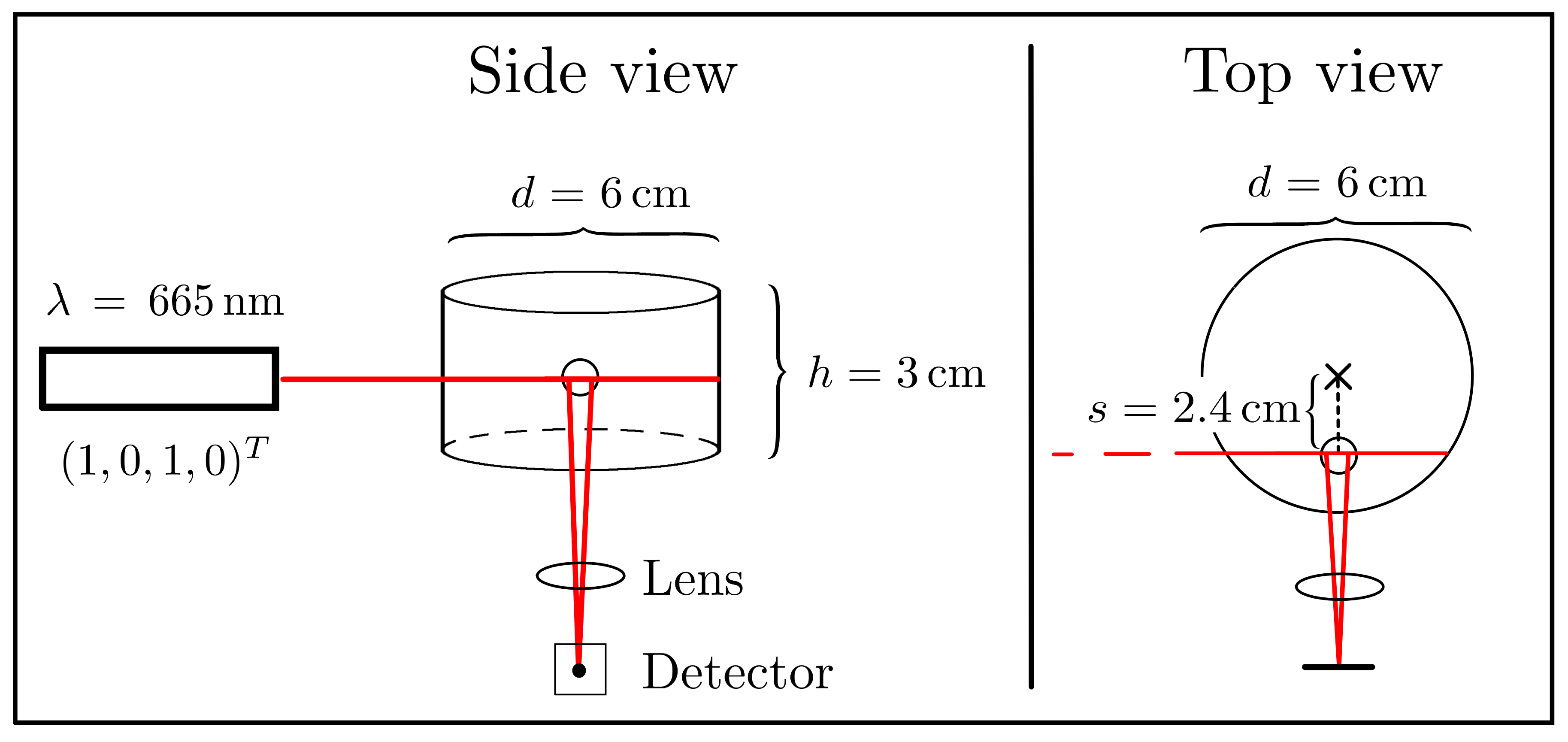} 
  \caption{Model setup for the radiative transfer simulations (not to scale), based on \cite{Groth2015}.}
 \label{System_sketch}
 \end{figure}

The nanodust is produced in a cylindrical plasma chamber by means of an RF driven capacitively coupled, parallel plate reactor at typical $\unit[10]{W}@\unit[13.56]{MHz}$ and an argon pressure of typical $\unit[20]{Pa}$. For details of the experimental setup see \cite{Tadsen2015}. Kinetic Mie ellipsometry is used to get information about size and density of the dust \cite{Groth2015}. A red laser ($\lambda=\unit[665]{nm}$) passes the cylindrical nanodust cloud at a distance of $\unit[2.4]{cm}$ from the symmetry axis (see Fig.~\ref{System_sketch}) to ensure a lower optical depth and to avoid passing the void in the center region. The scattered light is measured under an observing angle of $90^\circ$ with respect to the laser beam and within a circular area with a diameter of $\unit[0.5]{cm}$ at the position of the laser beam. The laser radiation is linearly polarized with an angle of $45^\circ$ with respect to the scattering plane before it enters the plasma chamber. 

The scattering process and the polarization state of the scattered light are described using the Stokes formalism with the Stokes vector $(I,Q,U,V)^{T}$  \mbox{\cite{Collett2012}}. The normalized Stokes vector $(q,u,v) = (Q/I_\text{p}, U/I_\text{p}, V/I_\text{p})$ can be determined from the measured scattered light,  where \mbox{$I_{\textrm{p}}=(Q^2+U^2+V^2)^{1/2}$} is the polarized fraction of the total intensity. The normalized Stokes vector $(q,u,v)$ is a function of the time-dependent particle radius $a(t)$ and the complex refractive index $N$.


\label{Sec_model_and_simu}
We apply the 3D polarized radiative transfer program \texttt{Mol3D} which is based on the Monte Carlo method \cite{Ober2015, Wolf1999, Wolf2003}. It has been developed for the simulation of observables of astrophysical objects which are dominated in their appearance by dust particles of nanometer to millimeter size. For the purpose of our study this program was adjusted to simulate the radiative transfer in a laboratory dusty plasma.

The radiation of the light source is represented by photon packages, each of which is characterized by its wavelength $\lambda$ and Stokes vector. To mimic the experimental setup, these photon packages are emitted from the light source towards the dust cloud where they can be scattered or absorbed by the dust material. 

\texttt{Mol3D} handles spherically symmetric absorbers and scatterers. The required optical properties, e.$\,$g. the absorption and scattering cross sections ($C_\text{abs}$, $C_\text{sca}$) and the scattering function, are calculated with an embedded Mie scattering routine \cite{Mie1908, BohrenHuffman1983, WolfVoshchinnikov04} on the basis of the complex refractive index $N$ and the particle radius $a$. The radiative transfer simulation is feasible in arbitrary dust density distributions, analytically or on a predefined three-dimensional density grid. The direct radiation of the light source as well as the thermal reemission of the dust can interact (absorption and scattering) with the dusty medium multiple times before leaving the model space. A detailed description of the applied scattering and absorption scheme is given in \cite{Wolf2003}. 


\begin{figure} 
  \includegraphics[trim=1.85cm 2.0cm 0.2cm 5.65cm, clip=true,width=1.0\linewidth]{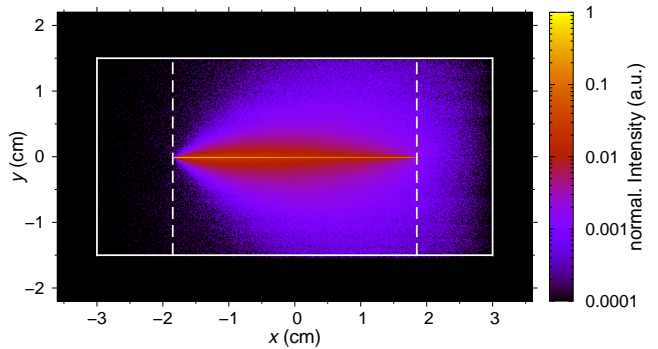} 
  \caption{Simulated intensity distribution of the scattered light for a number density scale parameter $\alpha=1.0$ and a dust grain radius $a=\unit[300]{nm}$. The cylinder is centered at the coordinate origin (right-handed system) and the polarized light source is located  at $(x,y,z)=(\unit[-3.2]{cm},0,\unit[2.4]{cm})$. The cylinder contours are shown as white lines. The initial path of the laser beam can be seen as horizontal line in the center of the image. The colour scale is logarithmic and normalized to unity.}
 \label{Result_Bild2D}
 \end{figure}

Our model consists of three components. These are the light source, the cylindrical dust cloud and the detector (Fig.~\ref{System_sketch}).

The light source is monochromatic ($\lambda=\unit[665]{nm}$) and directed solely towards the dust cloud. The initial Stokes vector is $(1,0,1,0)^{T}$ to take into account the polarization of the light source (linearly polarized $45^\circ$ to the scattering plane). 

The dust cloud is confined in a homogeneous cylinder of height $h=\unit[3]{cm}$ and diameter $d=\unit[6]{cm}$. The photons pass the cylinder perpendicular to the symmetry axis at a distance of $s=\unit[2.4]{cm}$ from it.  The number density of particles is constant and assumed to be $n=\unit[\alpha\times4.6\times10^{13}]{m^{-3}}$. For \mbox{$\alpha=1$} this corresponds to the experimental situation we want to compare with \cite{Tadsen2015}. We vary the number density scale parameter $\alpha$ between $0.06$ and $6$ to simulate dust clouds with different optical depths $\tau$. The dust particles are assumed to be spherical and monodisperse \cite{Denysenko2006, Greiner2012}. The grain radius $a$ is varied from $\unit[20]{nm}$ to $\unit[300]{nm}$ in steps of $\unit[5]{nm}$. The complex refractive index is taken from the experiment, $N=1.54 + 0.02\,\text{i}$ \cite{Groth2015}. 

We are interested in normalized intensities. Therefore, the distances between the laser and dust cloud as well as between the dust cloud and detector are irrelevant. In the experiment, the detector is located far enough from the plasma chamber to assume nearly parallel trajectories of the incident photons.  The Stokes vector $I$, $Q$, $U$ and $V$ are integrated over a circular area with diameter of $\unit[0.5]{cm}$ as depicted in Figure~\ref{System_sketch}. Subsequently, the normalized Stokes vector $(q,u,v)$ is determined. According to the experimental setup we assume that only photons directed within a narrow angular range ($\sim5^\circ$) towards the detector will be observed. 

The optical depth $\tau$ of the dust cloud is calculated along the path $w=\sqrt{d^2-4\,s^2}$ in the direction of the incident laser beam and amounts to
\begin{equation}
\tau=C_\text{ext}\,n\,w, 
\end{equation}
where $C_\text{ext}$ is the extinction cross-section of the dust particles. Thus, the optical depth $\tau$ is in direct proportion to the number density scale parameter $\alpha$.

A simulated 2D image of the scattered light intensity is shown in Figure~\ref{Result_Bild2D} for the number density scale parameter $\alpha=1.0$ and a dust grain radius $a=\unit[300]{nm}$. This image shows the spreading of the incoming photons over the whole \mbox{nano}dust cloud leading to multiple scattering.

\begin{figure} 
  \includegraphics[trim=2.05cm 2.0cm 1.0cm 3.4cm, clip=true,width=1.0\linewidth]{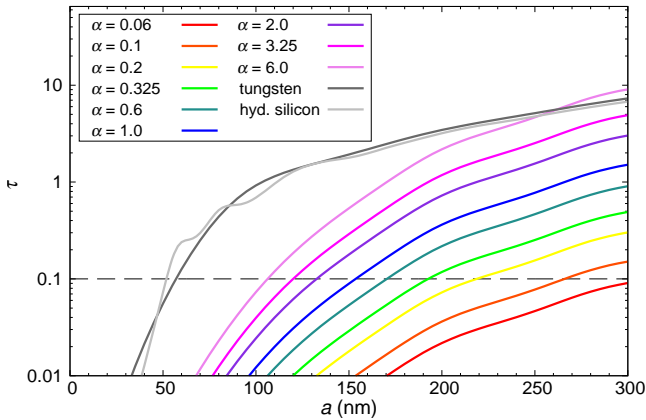} 
  \caption{Optical depth $\tau$ of the  nanoparticle cloud as a function of dust grain radius $a$ and number density scale parameter $\alpha$. For $\alpha=1.0$ the number density amounts to the experimental value $n=\unit[4.6\times10^{13}]{m^{-3}}$. A value of $\tau=0.1$ indicates the upper limit for which multiple scattering can be ignored, confirming the rule of thumb of van de Hulst \cite{Hulst1957}. The refractive index is $N=1.54+0.02\,\text{i}$. In addition, the optical depths at $\alpha=1.0$ for tungsten ($N=3.3+2.5\,\text{i}$) and hydrogenated silicon ($N=5.5+0.8\,\text{i}$) are shown.}
 \label{Result_Opt_depth}
 \end{figure}

\begin{figure} 
  \includegraphics[trim=1.8cm 2.0cm 1.0cm 3.25cm, clip=true,width=1.0\linewidth]{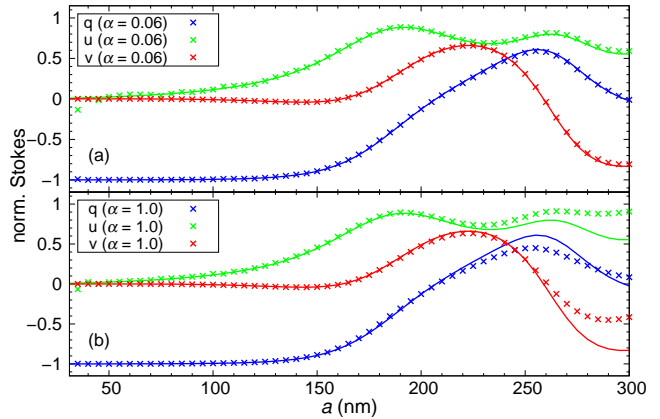} 
  \caption{Normalized Stokes vector $(q,u,v)$ as a function of the dust grain radius $a$ for the theoretical single scattering results calculated directly via Mie-theory (solid lines) and for $\alpha=0.06$ ((a), symbols) and $\alpha=1.0$ ((b), symbols). While  $\alpha=0.06$ shows no deviations, those are increased for the high-density case.}
 \label{Result_opt_thin}
 \end{figure}

The optical depth of the dust cloud with $\alpha=0.06$ is below $0.1$ for all grain sizes (Fig.~\ref{Result_Opt_depth}). Thus, multiple scattering can be ignored \cite{Hulst1957}. Consequently, the normalized Stokes parameter $(q,u,v)$ derived in this case show very good agreement with the single scattering results which are derived directly from Mie-theory (Fig.~\ref{Result_opt_thin}, (a)). With increased number density, the optical depth increases and multiple scattering becomes more and more important (e.$\,$g.~for $\alpha=1.0$, Fig.~\ref{Result_opt_thin}, (b)). As a result, the deviations between the single scattering results and the radiative transfer simulations grow with increasing optical depth.

Figure~\ref{Result_Opt_depth} also shows the optical depth for tungsten ($N=3.3+2.5\,\text{i}$, \cite{Bass2001}),  and hydrogenated silicon (\mbox{$N=5.5+0.8\,\text{i}$,} \cite{Hollenstein1994}).
For these materials, multiple scattering becomes important also for grain radii $a\lesssim\unit[50]{nm}$ for a number density scale parameter $\alpha=1.0$, emphasizing the relevance of radiative transfer simulations in systems with grains of a few tens of nanometer in size.

For comparison of our simulation results with the experimental measurements we use the ellipsometric angles \mbox{$\Psi\in [0,\pi/2]$} and $\Delta\in [-\pi,\pi]$ to describe the polarization state of the scattered light. The ellipsometric angles $\Psi$ and $\Delta$ present the intensity ratio and the relative phase of two perpendicular vibrating electromagnetic waves creating elliptically polarized light in the most general case. $\Psi$ and $\Delta$ can be directly calculated from the normalized Stokes vector.

\begin{figure*} 
  \includegraphics[trim=1.95cm 2.65cm 1.4cm 5.0cm, clip=true,width=1.0\linewidth]{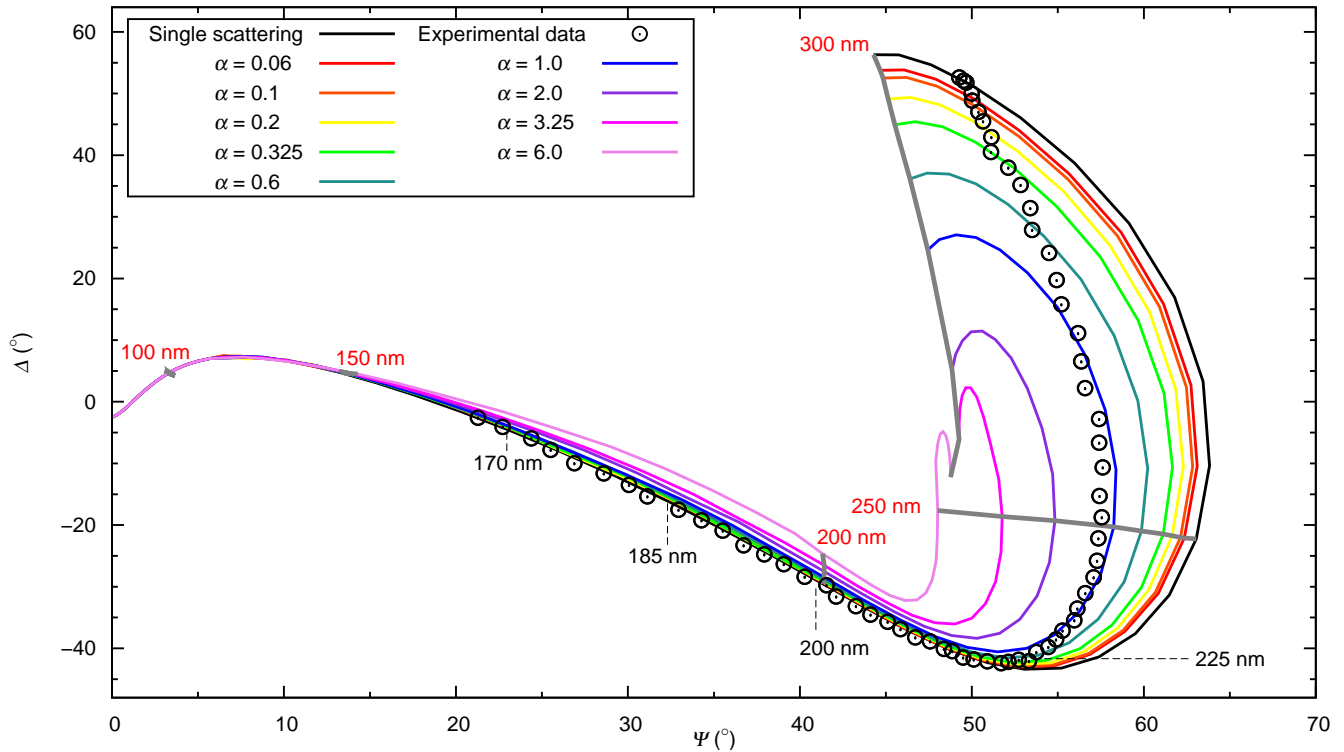} 
  \caption{Relation of the ellipsometric angles $\Psi$ and $\Delta$ for a-C:H particles as a function of the number density scale parameter $\alpha$ as well as the theoretical single scattering results calculated directly via Mie-theory and experimental data. Different dust grain radii are indicated \fk{for the simulations (red values) and the experimental data (black values). For the latter, the particle size is derived only for $a<\unit[226]{nm}$ \cite{Groth2015}.} The deviations between the radiative transfer simulations and the single scattering results increase with increasing optical depth, in particular for dust grain radii $a\gtrsim\unit[200]{nm}$.}
 \label{Result_Psi_delta}
 \end{figure*} 

We simulate the radiative transfer for nine different values of $\alpha$  between $0.06$ and $6.0$, calculate the ellipsometric angles $\Psi$ and $\Delta$ and compare them to the theoretical single scattering results calculated directly via Mie-theory (Fig.~\ref{Result_Psi_delta}). For a fixed value of $\alpha$, the $\Psi-\Delta$~relation has the shape of a pipe. The dust grain size increases along the pipe curve. The deviations between the radiative transfer simulations and the single scattering results increase with increasing optical depth, in particular for dust grain radii $a\gtrsim\unit[200]{nm}$. In addition, for a fixed grain radius the $\Psi-\Delta$~pairs are located on a nearly straight line in the $\Psi-\Delta$~diagram.

For comparison, the experimental measurements taken from \cite{Groth2015} are shown in Figure~\ref{Result_Psi_delta}. 
Up to a grain radius of $a=\unit[270]{nm}$ an excellent agreement between the experimental data and the simulation for $\alpha=1.0$ is found. Nevertheless, for larger grain radii the model results show significant deviations to the experimental data, \fk{where the number density decreases from $\unit[{\sim}5\times10^{13}]{m^{-3}}$ to $\unit[{\sim}0.25\times10^{13}]{m^{-3}}$ with increasing grain radius.} However, this has to be attributed to the simplicity of the chosen model, i.$\,$e., a constant density of the particle cloud. In the experiment, a thinning of the particle cloud during the growth is observed for larger particles. An advanced model for the dust number density as a function of the particle radius is required to take into account this particle leakage. 


We presented 3D polarized radiative transfer simulations to describe the light scattering in nanoparticle clouds. While optically thin systems ($\tau\lesssim0.1$) can be analyzed using a single scattering approach, the radiative transfer simulations consider multiple scattering on the dust particles, enabling the investigation of systems with higher optical depths. 

For experiments in an argon-acetylene plasma in an RF driven low pressure plasma, the change in the polarization state of the scattered light over time could be reproduced, even for larger particles, where the optical depth is not negligible. For the simulations presented here, we used a model which assumes a homogeneous cylindrical dust cloud and mono disperse particles. The incorporation of a size distribution of particles and realistic density profiles is straight forward.  For the presented experiments of a-C:H particle growth in a low temperature plasma multiple scattered light plays a significant role only for large particle sizes. The application of radiative transfer codes like \texttt{Mol3D} will help to develop high precision in-situ size diagnostics.  In combination with enhanced imaging Mie methods benchmarking and detailed comparison with simulation results of particle growth in low pressure plasmas \cite{Agarwal2014} will be possible. For the Mie diagnostic of atmospheric pressure plasma jets used to produce nanoparticles \cite{Nozaki2007, Kortshagen2016}, it is essential  to use radiative transfer codes for the interpretation of  the scattered light, due to the high density of the particles and the strong inhomogeneity of the particle flow.
  
In the future, the dust properties and the density structure of the cloud can be derived from radiation which was scattered within the dust particle cloud. 
\newline~\newline

We would like to thank Gesa Bertrang for her contribution in the initial phase of this project. This work was supported by the Deutsche Forschungsgemeinschaft DFG in the framework of \mbox{SFB-TR24} Greifswald Kiel, Projects A2 and A5, and under the contract WO 857/15-1.


\end{document}